\documentclass[12pt]{article}
\usepackage[centertags]{amsmath}
\usepackage{amsfonts}
\usepackage{amssymb}
\usepackage{amsthm} 
\usepackage{newlfont}
\usepackage{epsfig}
\usepackage{amscd}
\newcommand{\beq}{\begin{equation}}
\newcommand{\eeq}{\end{equation}}
\newcommand{\ba}{\begin{array}}
\newcommand{\ea}{\end{array}}
\newcommand{\bea}{\begin{eqnarray}}
\newcommand{\eea}{\end{eqnarray}}

\begin{document}

\begin{center}
{\large \sc \bf Partially integrable  generalizations of 
classical integrable models by combination of  characteristics method and Hopf-Cole transformation }

\vskip 15pt

%{\large P. M. Santini$^{1,\S}$ and A. I. Zenchuk$^{2,\S}$ }
{\large  A. I. Zenchuk }

\vskip 8pt

%{\it
%$^1$ Dipartimento di Fisica, Universit\`a di Roma "La Sapienza" and\\
%Istituto Nazionale di Fisica Nucleare, Sezione di Roma 1\\
%Piazz.le Aldo Moro 2, I-00185 Roma, Italy}

\smallskip

{\it  Institute of Chemical Physics, RAS,
Acad. Semenov av., 1
Chernogolovka,
Moscow region
142432,
Russia}

\smallskip

\vskip 5pt

e-mail:  {\tt zenchuk@itp.ac.ru }

\vskip 5pt

{\today}

\end{center}

\begin{abstract}
We represent an integration algorithm  combining the  characteristics method and Hopf-Cole transformation. This algorithm
allows one to partially integrate a large class of multidimensional systems of  nonlinear Partial Differential Equations (PDEs).
 A specific generalization of the equation describing the dynamics of two-dimensional viscous fluid and 
 a generalization of the Korteweg-de Vries equation are examples of such systems. The richness of available solution space for derived nonlinear PDEs 
is discussed. 
\end{abstract}

%%%%%%%%%%%%%%%%
\section{Introduction}
A possible way to construct new classes of (partially) integrable nonlinear multidimensional Partial Differential Equations (PDEs) is combining features of different classical integration algorithms. Thus,  the multidimensional 
nonlinear PDEs having features of PDEs linearizable ($C$-integrable \cite{Calogero}) by the Hopf-Cole transformation \cite{Hopf,Cole} and integrable by the 
 Inverse Spectral Transform Method (ISTM)  \cite{ZSh1,ZSh2,ZM,BM} ($S$-integrable)  have been derived in refs.\cite{Z1,Z3}. A  combination of the characteristics method \cite{KG} with the Hopf-Cole transformation is proposed in refs.\cite{SZ2,Z3}. The characteristics method was also combined with the  ISTM in refs. \cite{Z2,Z3} and with  the commuting vector fields \cite{MS1,MS2} in ref. \cite{Z4}.
There are also   several generalizations of the characteristics method \cite{T,DN,T2,F,K} integrating  certain classes of systems of multidimensional  nonlinear PDEs. In addition, there is a relation \cite{SZ} between the nonlinear PDEs linearizable by the Hopf-Cole transformation and integrable by the ISTM. 

The algorithm proposed in this paper allows one to partially integrate  a new class of 
multi-dimensional nonlinear PDEs using the features of the characteristics method and Hopf-Cole transformation. Unlike the nonlinear PDEs derived in  refs.\cite{SZ2,Z3}, a new type of nonlinear PDEs involves additional arbitrary functions of fields and their derivatives. 
 Note that the term  "partially integrable PDEs"  has two different meanings. First, the PDE is 
partially integrable if the available   solution space is not full \cite{ZS}.
Second, the PDE is partially integrable if its  solution space is described in terms of the  lower dimensional PDEs \cite{Zen}. Our nonlinear  PDEs are partially integrable in both senses.

The general result is covered by the following theorem.

{\bf Theorem.}
Let $W(x,t)$  be a solution to the  algebraic system 
 \begin{eqnarray}\label{W1}
&&
W(x,t) F(u(x,t),x,t) =\eta(u(x,t)),\\
 \label{W2}
&&
W(x,t) F_{u^{(n)}}(u(x,t),x,t) -\eta_{u^{(n)}}(u(x,t)) =\eta^{(n)}(x,t),\;\;n=1,\dots,K,
\\\label{det}
&&
\det W\neq 0,\\\label{tx}
&&
t=\{t_1,\dots,t_M\},\;\;x=\{x_1,\dots,x_N\}, \;\;
u=\{u^{(1)},\dots,u^{(K)}\},
\end{eqnarray}
where  $W$ is the $N_1\times N_1$ matrix function of  independent variables $x$ and $t$; $F$,
 $\eta$ and $\eta^{(n)}$ are the $N_1\times N_2$ ($N_1>N_2$) matrix functions of arguments with
 \begin{eqnarray}
\label{NN}
N_1=N_2 (K+1).
\end{eqnarray}
Let $\eta$ and $\eta^{(n)}$ be arbitrary functions, while  $F$  satisfy the following system of compatible linear PDEs:
\begin{eqnarray}\label{Ft}
F_{t_m} - \sum_{j=1}^M F_{x_j} V^{(mj)}(u)=F \Gamma^{(m)}(u),\;\;m=1,\dots,M,
\end{eqnarray}
where $ V^{(mj)}$ and $\Gamma^{(m)}$, are arbitrary $N_2\times N_2$ diagonal matrix functions of $u$.
Then elements of the matrix function $W$  satisfy the following system of nonlinear  equations:
\begin{eqnarray}\label{nlin}
&&
W_{t_m} W^{-1} \eta(u) -\sum_{j=1}^N W_{x_j} W^{-1} \eta(u) V^{(mj)}(u) + \\\nonumber
&&
\sum_{i=1}^K  \Big(u^{(i)}_{t_m} -
\sum_{j=1}^N u^{(i)}_{x_j} V^{(mj)}(u)\Big)\eta^{(i)}(x,t)+ \eta(u) \Gamma^{(m)}(u)=0,
\end{eqnarray}
where $u^{(i)}$ are arbitrary functions of $x$ and $t$.

{\it Proof.}
First, we  differentiate eq.(\ref{W1}) with respect to $t_m$ and $x_j$:
\begin{eqnarray}\label{Wt}
&&
E^{(t_m)}:=W_{t_m} F + W F_{t_{m}} + W\sum_{i=1}^K F_{u^{(i)}} u^{(i)}_{t_m} =
\sum_{i=1}^K\eta_{u^{(i)}} u^{(i)}_{t_m},\\\label{Wx}
&&
E^{(x_n)}:=W_{x_n} F + W F_{x_n} + W\sum_{i=1}^K F_{u^{(i)}} u^{(i)}_{x_n} =
\sum_{i=1}^K\eta_{u^{(i)}} u^{(i)}_{x_n}.
\end{eqnarray}
Consider the  combination
$E^{(t_m)}-\sum_{j=1}^N E^{(x_j)} V^{(mj)}$ and take into account eqs.(\ref{W2}) and (\ref{Ft}). We obtain:
\begin{eqnarray}\label{W_int}
W_{t_m} F -\sum_{j=1}^N W_{x_j} F V^{(mj)} + \sum_{i=1}^K \eta^{(i)} \Big(u^{(i)}_{t_m} - \sum_{j=1}^N u^{(i)}_{x_j} V^{(mj)}\Big) +
W F \Gamma^{(m)}(u)= 0.
\end{eqnarray}
Since $W$ is invertible, from eq.(\ref{W1}) we have: $F= W^{-1}\eta$. Substituting it into eq.(\ref{W_int}) we obtain eq.(\ref{nlin}).
$\Box$

{\bf Remark 1.} If $W=1$, $\eta=0$, $N_1=1$, $N_2=K$ and $\det \{(F_{1i})_{u^{(j)}}: i,j=1,\dots K\}\neq 0$, then eq.(\ref{W2}) may be disregarded. Herewith eq.(\ref{W1}) yields the  usual characteristics method \cite{KG} and eq.(\ref{nlin}) must be replaced with the system
\begin{eqnarray}
u^{(i)}_{t_m} - \sum_{j=1}^N u^{(i)}_{x_j} V^{(mj)} =0,\;\;i=1,\dots,K,\;\;m=1,2,\dots.
\end{eqnarray}
If $N_1=N_2$, $\eta = F_{x_1}$ and $\det F\neq 0$, then $W=F_{x_1} F^{-1}$ so that  we have  the matrix Hopf-Cole transformation. To derive the nonlinear PDE for the matrix $W$ in this case we have to replace eqs.(\ref{Ft}) by another  system for $F$. For instance, if
\begin{eqnarray}
F_{t_m} = \partial_x^m F + F B^{(m)} - B^{(m)} F,\;\;m=1,2,\dots,
\end{eqnarray}
(where $B^{(m)}$ are diagonal constant matrices)
then $W$ satisfies the matrix B\"urgers hierarchy \cite{BLR}.

{\bf Remark 2.}
If $N_2=1$, i.e.
$K=N_1-1$, then eqs.(\ref{W1}) and (\ref{W2}) may be written as the following single matrix equation:
\begin{eqnarray}
W \hat F = \hat \eta,
\end{eqnarray}
where $\hat F$ and $\hat \eta$ are   block-row matrices
\begin{eqnarray}
\hat F= [F \;\;F_{u^{(1)}} \;\dots \; F_{u^{(N_1-1)}} ],\;\;
\hat\eta=[\eta \;\; \eta^{(1)}+\eta_{u^{(1)}}\; \dots \; \eta^{(N_1-1)}+\eta_{u^{(N_1-1)}}]
.
\end{eqnarray}
To satisfy the condition  (\ref{det}) we require
$\det \;\hat F \neq0$ and $ \det \hat \eta \neq 0$.
Then 
\begin{eqnarray}\label{hatW}
W  = \hat \eta \hat F^{-1}.
\end{eqnarray}

Emphasize, that $\eta(u)$, $\eta^{(i)}(x,t)$ and $V^{(mj)}(u)$ are arbitrary functions of arguments with appropriate matrix dimensions.

Note that functions $u^{(i)}$ and $\eta^{(i)}$ ($i=1,\dots,K$)  are arbitrary functions of $x$ and $t$ in 
eq.(\ref{nlin}). Consequently, we have to 
add  relations among $u^{(j)}$, $\eta^{(i)}$ and elements of $W$. Further, each matrix  equation with fixed $m$ 
in the  system (\ref{nlin})  involves 
 $N_1 N_2$ scalar equations for $N_1^2$ scalar fields $W_{ij}$ ($i,j=1, \dots, N_1$).  Thus, the  complete   system  consists of  
several equations (\ref{nlin}) with different $m=1,\dots,T$,  so that
$N_1 N_2 T = N_1^2$. 
Consequently, the nonlinear system is $N+T$-dimensional.

\section{Nonlinear PDEs corresponding to  $N_1=2$, $N_2=1$ }

Consider the case $N_1=2$, $N_2=1$, $T=2$  and denote $u^{(1)}\equiv u$. 
Then the ($N+2$)-dimensional system of nonlinear PDEs (\ref{nlin}) reads:
\begin{eqnarray}\label{nlin3}
&&
E^{(m)}_\alpha : =A^{w;1} \left( (W_{\alpha 1})_{t_m} - \sum_{j=1}^N (W_{\alpha 1})_{x_j} V^{(mj)}_1(u)\right) + \\\nonumber
&&
A^{w;2} \left( (W_{\alpha 2})_{t_m} - \sum_{j=1}^N (W_{\alpha 2})_{x_j} V^{(mj)}_1(u)\right) +\\\nonumber
&&
A^{(v)} \eta^{(1)}_{\alpha 1}(W,u) \left( u_{t_m} - \sum_{j=1}^N u_{x_j} V^{(mj)}_1(u)\right) +A^{(v)}\eta_{\alpha 1}\Gamma^{(m)}_1=0,
\;\;\alpha=1,2,\;\;m=1,2.
\end{eqnarray}
where
\begin{eqnarray}
&&
A^{w;1}= W_{22}\eta_{11}(u) - W_{12} \eta_{21}(u),\;\;A^{w;2}= W_{11}\eta_{21}(u) - W_{21} \eta_{11}(u),\\\nonumber
&&
A^{v}= (W_{11}W_{22}- W_{12} W_{21}) .
\end{eqnarray}
Let $\eta$ be independent on $u$ for the sake of simplicity,
\begin{eqnarray}\label{Wuv}
u=W_{11},\;\;v=W_{12}, \;\;p=W_{21},\;\;q=W_{22}.
%\;\;A^{(v)} (\eta^{(1)}_{11} \eta_{21} - \eta^{(1)}_{21}\eta_{11} )= G(x,u,v). 
\end{eqnarray}
Then, introducing the operators $L^{(m)}$ acting on an arbitrary function $h(x,t)$ by the formula  
\begin{eqnarray}\label{L}
L^{(m)}(h) = h_{t_m} - \sum_{j=1}^N  h_{x_j} V^{(mj)}_1(u),
\end{eqnarray}
we write  eq.(\ref{nlin3}) as
%\begin{eqnarray}\label{nlin4_1}
%&&
%u_{t_m} - \sum_k u_{x_k} V^{(mk)}_1(u) 
%G_{11} \left( v_{t_m} - \sum_k v_{x_k} V^{(mk)}_1(u)\right) +
%G_{12}\Gamma^{(m)}_1=0,\;\;m=1,2,\\\label{nlin4_2}&&
%p_{t_m} - \sum_k p_{x_k} V^{(mk)}_1(u) + 
%\frac{A^{(w;2)}}{A^{(W;1)}} \left( q_{t_m} - \sum_k q_{x_k} V^{(mk)}_1(u)\right)  +
%G_2\Gamma^{(m)}_1=0,\;\;m=1,2\\\label{GG1}
%&&
%G_1 = \frac{A^{(w;2)}}{A^{(v)}\eta^{(1)}_{11} + A^{(w;1)}}=\frac{u\eta_{21} - p \eta_{11}}{\eta^{(1)}_{11}(pv-uq) +(q\eta_{11}-v\eta_{21})},
%\\\label{GG2}
%&&
%G_2 =
%\frac{A^{(v)} \eta_{11} }{A^{(v)}\eta^{(1)}_{11} + A^{(w;1)}}=\frac{\eta_{11}(pv-uq)}{\eta^{(1)}_{11}(pv-uq) +(q\eta_{11}-v\eta_{21})}.
%\end{eqnarray}
\begin{eqnarray}\label{nlin4_1}
&&
L^{(m)}(u) +
G_{11} L^{(m)}(v)  +
G_{12}\Gamma^{(m)}_1=0,\;\;m=1,2,\\\label{nlin4_2}&&
L^{(m)}(p) + G_{21}  L^{(m)}(q) + G_{22} L^{(m)}(v) + G_{23}\Gamma^{(m)}_1 = 0,\;\;m=1,2\\\label{GG1}
&&
G_{11} = \frac{A^{(w;2)}}{A^{(v)}\eta^{(1)}_{11} + A^{(w;1)}},
%=\frac{u\eta_{21} - p \eta_{11}}{\eta^{(1)}_{11}(pv-uq) +(q\eta_{11}-v\eta_{21})},
\\\label{GG2}
&&
G_{12} =
\frac{A^{(v)} \eta_{11} }{A^{(v)}\eta^{(1)}_{11} + A^{(w;1)}},
%=\frac{\eta_{11}(pv-uq)}{\eta^{(1)}_{11}(pv-uq) +(q\eta_{11}-v\eta_{21})}.
\\\label{GG3}
&&
G_{21}=\frac{A^{(w;2)}}{A^{(w;1)}} ,
\\\label{GG4}
&&
G_{22}=-\frac{A^{(v)} A^{(w;2)} \eta^{(1)}_{21}}{A^{(w;1)} ( A^{(w;1)} + A^{(v)} \eta^{(1)}_{11})} ,\\\label{GG5}
&&
G_{23}=\frac{\eta_{21} A^{(v)}}{A^{(w;1)}} - \frac{\eta_{11} \eta^{(1)}_{21} (A^{(v)})^2 }{A^{(w;1)} ( A^{(w;1)} + A^{(v)} \eta^{(1)}_{11})}
\end{eqnarray}
where 
\begin{eqnarray}
\label{AA}
A^{(w;1)} = q \eta_{11} - v \eta_{21},\;\; A^{(w;2)} = u \eta_{21} - p \eta_{11},\;\;
A^{(v)} = uq-pv.
\end{eqnarray}
Owing to the arbitrary functions $\eta^{(1)}_{i1}(x,t)$ ($i=1,2$), coefficients $G_{12}$ and $G_{23}$ may be arbitrary functions of $x$ and $t$. Consequently, we may impose the following relations
\begin{eqnarray}
\label{GGf}
G_{12}= f_1(U),\;\;G_{23}= f_2(U)
\end{eqnarray}
where list  of arguments $U$ in arbitrary functions $f_i$ ($i=1,2$) involves the  
fields  $u$, $v$, $p$, $q$  and  their derivatives: $U=\{u,v,p,q, u_x,v_x,p_x,q_x,u_t,v_t,p_t,q_t,\dots\}$.
Substituting $G_{12}$ and $G_{23}$  from eqs.(\ref{GG2}) and (\ref{GG5}) into eqs. (\ref{GGf}) and solving them
 for $\eta^{(1)}_{i1}$, $i=1,2$, we obtain:
\begin{eqnarray}\label{eta111}
\eta^{(1)}_{i1} =\frac{A^{(v)} \eta_{i1} - A^{(w;1)} f_i }{A^{(v)} f_1},\;\;i=1,2.
%\frac{\eta_{11}(f q + q u - pv) -fv \eta_{21}}{f(qu -pv)}
\end{eqnarray}
Now eqs.(\ref{nlin4_1}) and  (\ref{nlin4_2})  read as 
\begin{eqnarray}\label{nlin4_1_2}
&&
L^{(m)}(u) +
G_{11} L^{(m)}(v)  +
f_1\Gamma^{(m)}_1=0,\;\;m=1,2,\\\label{nlin4_2_2}&&
L^{(m)}(p) + G_{21}  L^{(m)}(q) + G_{22} L^{(m)}(v) + f_2 \Gamma^{(m)}_1 = 0,\;\;m=1,2,
\end{eqnarray}
%\begin{eqnarray}
%\eta^{(1)}_{11} =
%\frac{\eta_{11}}{f} + F_{11} 
%\end{eqnarray}
where, in virtue of eqs.(\ref{eta111}), $G_{ij}$  read
\begin{eqnarray}
&&
G_{11}=\frac{A^{(w;2)} f_1}{A^{(v)} \eta_{11}}=\frac{f_1(u\eta_{21}-p\eta_{11})}{\eta_{11}( u q-pv)},\\\nonumber
&&
G_{21}=\frac{A^{(w;2)}}{A^{(w;1)}}=\frac{u\eta_{21}-p\eta_{11}}{q \eta_{11} - v \eta_{21}},\\\nonumber 
&&
G_{22} =\frac{A^{(w;2)} (A^{(w;1)} f_2 - A^{(v)} \eta_{21})}{A^{(v)} A^{(w;1)} \eta_{11}}=
\frac{(u\eta_{21} - p\eta_{11})(v\eta_{21}(p-f_2) - q(u\eta_{21}-\eta_{11} f_2))}{\eta_{11}(qu-pv)(q\eta_{11} - v \eta_{21})}.
\end{eqnarray}
Eqs.(\ref{nlin4_1_2}) and (\ref{nlin4_1_2}) must be considered as the ultimate
 general ($N+2$)-dimensional  system of nonlinear  PDEs  in this example. 
%%%%%%%%%%%%%%%
\subsection{Reduction of the system (\ref{nlin4_1_2}, \ref{nlin4_2_2}) to the  single PDE}
There is a particular reduction of the system (\ref{nlin4_1_2}, \ref{nlin4_2_2}) leading to the single nonlinear PDE
for the function $u$.  Namely, if the second term in the system (\ref{nlin4_1_2}) is negligible, then this system reads
\begin{eqnarray}\label{uu}
&&
u_{t_m} - \sum_{j=1}^N u_{x_j} V^{(mj)}_1(u) +
f_1 \Gamma^{(m)}_1=0,\;\;m=1,2,
\end{eqnarray}
i.e. $u$ satisfies the system of two commuting flows. In this case, it is reasonable to take $f_1$ as an arbitrary function of $u$ and its derivatives, while  eq.(\ref{nlin4_2_2}) may be disregarded. 
Of course, commutativity condition  imposes additional differential relation on the field $u$.
To avoid this relation we have to disregard eqs.(\ref{uu}) with $m=2$, keeping the only equation 
\begin{eqnarray}\label{uu1}
&&
u_{t_1} - \sum_{j=1}^N u_{x_j} V^{(1j)}_1(u) +
f_1 \Gamma^{(1)}_1=0.
\end{eqnarray}

This reduction may be realized by means of the multi-scale expansion with
\begin{eqnarray}\label{varepsilon}
u\sim v \sim p \sim q \sim \varepsilon^{(u)},\;\;
V^{(mk)}\sim\varepsilon^{(V)},
 \;\;\partial_{x_k} \sim \varepsilon^{(x)}, 
\;\;\partial_{t_m}\sim\varepsilon^{(x)} \varepsilon^{(V)},\;\;
f_1\sim\varepsilon^{(u)}\varepsilon^{(x)} \varepsilon^{(V)},
\end{eqnarray}
where all $\varepsilon$'s are small positive parameters ($\ll 1$).
In this case $G_{11}\sim \varepsilon^{(x)}  \varepsilon^{(V)}$. Consequently,  the second term in eq.(\ref{nlin4_1_2}) is of the order 
$ \varepsilon^{(u)}(\varepsilon^{(x)} \varepsilon^{(V)})^2$, while all other terms are of the order 
$ \varepsilon^{(u)}\varepsilon^{(x)} \varepsilon^{(V)}$. Thus, the leading
$ \varepsilon^{(u)}\varepsilon^{(x)} \varepsilon^{(V)}$-order  terms yield eq.(\ref{uu}).

\paragraph{Example 1: three-dimensional viscous fluid flow.} Let $N=3$,  $V^{(1k)}(u)=u$, $f=-\frac{\nu\Delta u}{\Gamma^{(1)}_1}$, ($\Delta=\sum_{i=1}^3 \partial_{x_i}$ 
is the three-dimensional Laplasian).
Then eq.(\ref{uu1}) yields a special case of three-dimensional viscous fluid flow with the constant kinematic viscosity  $\nu$
\begin{eqnarray}\label{uu1ex1}
&&
u_{t_1} - \sum_{j=1}^3 u_{x_j} u -
\nu\Delta u=0.
\end{eqnarray}
A two-dimensional ($M=2$) version of this equation 
was derived in \cite{Zen} using a different algorithm.

\paragraph{Example 2: multidimensional generalization of the Korteweg-de Vries equation (KdV).}
Let $V^{(11)}=u$, $V^{(1j)}=0$ ($j>1$), $f=\frac{ \sum_{i,j,k=1}^N a_{ijk} u_{x_ix_jx_k}}{\Gamma^{(1)}_1}$, $a_{ijk}=const$.
Then eq.(\ref{uu1}) yields an $(N+1)$-dimensional version of KdV:
\begin{eqnarray}\label{uu1ex2}
&&
u_{t_1} - u_{x_1} u +
\sum_{i,j,k=1}^N a_{ijk} u_{x_ix_jx_k}=0.
\end{eqnarray}

%%%%%%%%%%%%%
\subsection{Solution space to the nonlinear system (\ref{nlin4_1_2},\ref{nlin4_2_2})}
To obtain solutions to the derived system of nonlinear PDEs (\ref{nlin4_1_2},\ref{nlin4_2_2}) we, first, 
write expressions for the elements of the matrix function $W$ given by  the system  (\ref{hatW}) with $N_1=2$:
\begin{eqnarray}\label{Wsol}
W_{i1}=\frac{\eta^{(1)}_{i1} F_{21}-\eta_{i1} 
(F_{21})_u}{F_{21} (F_{11})_u-F_{11} (F_{21})_u},\;\;
W_{i2}=-\frac{\eta^{(1)}_{i1} F_{11}-\eta_{i1} 
(F_{11})_u}{F_{21} (F_{11})_u-F_{11} (F_{21})_u},
\end{eqnarray}
where $F_{\alpha 1}$ ($\alpha=1,2$) are 
solutions to eq.(\ref{Ft}):
\begin{eqnarray}
F_{\alpha1}= e^{\sum_{m=1}^M \Gamma^{(m)}_1 t_m} {\cal{F}}_{\alpha 1}
\left(x_1+ \sum_{m=1}^M t_m V^{(m1)}_1(u), \dots, 
x_N + \sum_{m=1}^M t_m V^{(mN)}_1(u)\right).
\end{eqnarray}
Substituting $W_{ij}$ from eqs.(\ref{Wsol}) and $\eta^{(1)}_{i1}$ ($i=1,2$) from eq.(\ref{eta111}) into 
the system (\ref{Wuv}) we obtain
\begin{eqnarray}\label{Luex11}
u&=&\frac{\eta_{11} F_{21}- f_1 F_{11} F_{21} -  f_1 \eta_{11} (F_{21})_u}
{f_1(F_{21} (F_{11})_u-F_{11} (F_{21})_u)},\\\label{Luex21}
p&=&\frac{\eta_{21} F_{21}- f_2 F_{11} F_{21} -  f_1 \eta_{21} (F_{21})_u}
{f_1(F_{21} (F_{11})_u-F_{11} (F_{21})_u)},\\\label{Luex12}
v&=&\frac{-\eta_{11} F_{11} + f_1 F_{11}^2 + f_1 \eta_{11} (F_{11})_u}
{f_1(F_{21} (F_{11})_u-F_{11} (F_{21})_u)},\\\label{Luex22}
q&=&\frac{-\eta_{21} F_{11} + f_2 F_{11}^2 + f_1 \eta_{21} (F_{11})_u}
{f_1(F_{21} (F_{11})_u-F_{11} (F_{21})_u)}.
\end{eqnarray}
If $f_i$ depend on all fields and their derivatives, then the system (\ref{Luex11}) is a lower dimensional system of PDEs for the fields $u,v,p,q$ (dimensionality of this PDE must be less then the dimensionality of  the system 
(\ref{nlin4_1_2},\ref{nlin4_2_2})). Derivatives in eqs.(\ref{Luex11}-\ref{Luex22}) appear due to the functions $f_i$ ($i=1,2$), 
which  may depend on the partial derivatives of the fields $u$, $v$, $p$ and $q$.
In particular, if $f_i$ ($i=1,2$) are the function of single field  $u$ and its derivatives, then only eq.(\ref{Luex11})  is a 
PDE for the function $u$, while 
eqs.(\ref{Luex21}-\ref{Luex22})   are  the non-differential equations for the fields  $v$, $p$ and $q$. 

Now we briefly characterize the 
solutions $u$, $v$, $p$ and $q$ to the system (\ref{nlin4_1_2},\ref{nlin4_2_2}), which are implicitly given   by eqs.(\ref{Luex11}-\ref{Luex22}). 
As was noted above, the system (\ref{Luex11}-\ref{Luex22})   is a system of  PDEs  for the fields $u$,
$v$, $p$, $q$  whose dimensionality is less then $N+2$, 
i.e. we do not completely integrate the original system of PDEs  (\ref{nlin4_1_2},\ref{nlin4_2_2}), but reduce its dimensionality. Further,  eqs.(\ref{Luex11}-\ref{Luex22}) involve two arbitrary functions  ${\cal{F}}_{i1}$, $i=1,2$,  of $N$ variables. 
For this reason, the available  solution space is not full (for the  fullness, the solution space to the  ($N+2$)-dimensional system of four  PDEs 
 must involve four 
arbitrary functions of $N+1$ variables). Thus the system (\ref{nlin4_1_2},\ref{nlin4_2_2})
 is partially integrable by our algorithm. Since all fields  
are given implicitly by eqs.(\ref{Luex11}-\ref{Luex22}), they describe the  wave breaking in general.

%%%%%%%%%%%%%%%%%
\section{Conclusions}

The algorithm proposed in this paper allows one to solve a new class of 
multi-dimensional nonlinear PDEs using the features of the characteristics method and Hopf-Cole transformation. 
 The principal 
%difference between the algorithm  proposed in ref.\cite{SZ2} and 
novelty of this algorithm is the presence  of arbitrary 
functions of fields
 in eq.(\ref{nlin}), i.e.  functions $\eta, \eta^{(i)}$ and $\Gamma^{(m)}$. Thus, this algorithm possesses
 new features which do not appear in 
both method of characteristics and method of direct linearization by the Hopf-Cole transformation. 
The derived  nonlinear PDEs are partially integrable since, first, one has to solve a lower dimensional PDE in order to construct  solutions and, second, the available solution space is not full.  The important question is whether the multi-scale expansion (\ref{varepsilon}) may be properly implemented  in the proposed algorithm so that eq.(\ref{uu1}) would be exactly solvable. 

It is important that,
among the  derived nonlinear PDEs, there are such equations that may be considered as multidimensional 
generalizations of known integrable models. 
 The physical applications of these generalizations  become obvious provided that some terms are negligible, 
see  eq.(\ref{uu1})  reduced from  the system (\ref{nlin4_1_2},\ref{nlin4_2_2}).

This work is supported by the RFBR grant  10-01-00787  and by the Program for Support of Leading Scientific Schools 
(grant No.6170.2012.2).

%%%%%%%%%%%%%%%%%%%%%%%%%%%%%%%%%%
%%%%%%%%%%%%%%%%%%%%%%%%%%%%%%%%%


\begin{thebibliography}{99}
%%%%%%%%%%%%%%%%%%%%%%%%%%%%%%%%%%
%%%%%%%%%%%%%%%%%%%%%%%%%%%%%%%%%%

\bibitem{Calogero}
F. Calogero  in ``What is Integrability'' ed.  V. E. Zakharov (Berlin: Springer), (1990) p.1


\bibitem{Hopf}
E. Hopf, Commun. Pure Appl. Math. {\bf 3} (1950) 201

\bibitem{Cole}
J.D. Cole, Q. Appl. Math. {\bf 9} (1951) 225

\bibitem{ZSh1}
V.E. Zakharov  and A.B. Shabat,  Funct. Anal. Appl. {\bf 8} (1974) 43

\bibitem{ZSh2}
V.E. Zakharov and A.B. Shabat,   Funct. Anal. Appl. {\bf 13} (1979) 13
 
\bibitem{ZM}
V.E. Zakharov  and S.V. Manakov,  Funct. Anal. Appl. {\bf 19} (1985) 11
 
\bibitem{BM}
L.V. Bogdanov  and S.V. Manakov, J. Phys. A: Math. Gen. {\bf 21} (1988) L537


\bibitem{Z1}
A.I. Zenchuk,
%{\it Unified dressing method for $C$- and $S$-integrable
%hierarchies.
%Particular example of (3+1) - dimensional $n$-wave equation
%}, 
 J.Physics A: Math.Gen. {\bf 37}, (2004) 6557 

\bibitem{Z3}
A.I. Zenchuk,
%{\it On the relations among nonlinear PDEs with different integrability properties: $C-S$-, $Ch-C$- and $Ch-S$-integrable systems}, 
in "Progress in Mathematical Physics Research", Nova Science Publishers, Inc., (2009) 605
%-619 


\bibitem{KG}
R. Kurant and D. Gilbert, {\it Methods of Mathematical Physics 2} (Wiley, New York, 1962)


\bibitem{SZ2}
 A.I. Zenchuk and P.M. Santini,
%{\it Dressing method based on homogeneous 
%Fredholm equation: quasilinear PDEs
%  in multidimensions},  
J.Phys.A:Math.Theor, {\bf 40}  
  (2007)  6147
 

\bibitem{Z2}
 A.I.Zenchuk,
% {\it Combination of  inverse spectral transform and 
% method of characteristics: deformed Pohlmeyer equation}, 
 proceedings to NEEDS'07, Journal of Nonlinear Mathematical Physics, {\bf 15}, Suppl. 3 (2008) 437
%-448, arXiv:0709.1196 [nlin.SI]\\

\bibitem{MS1}
S.V. Manakov and P.M. Santini, Phys. Lett. A {\bf 359}, (2006) 613

\bibitem{MS2}
S.V. Manakov and P.M. Santini, JETP Letters, {\bf 83}, No 10, (2006) 462


\bibitem{Z4}
A.I. Zenchuk, J.Math.Phys. {\bf 50} (2009) 063505 


\bibitem{T}
S.P. Tsarev, Sov. Math. Dokl. 31 (3) (1985) 488.

\bibitem{DN}
B.A. Dubrovin, S.P. Novikov, Russian Math. Surveys {\bf 44} (6) (1989) 35.

\bibitem{T2}
S.P. Tsarev, Math. USSR Izv.{\bf 37} (1991) 397.

\bibitem{F}
E.V. Ferapontov, Teor. Mat. Fiz. {\bf 99} (1994) 257.

\bibitem{K}
B.A. Kupershmidt, Lett. Math. Phys. {\bf 76} (2006) 1.



\bibitem{SZ}
A.I. Zenchuk and P.M. Santini,
J. Phys. A: Math. Theor. {\bf 41} (2008) 185209 

\bibitem{ZS}
A.I. Zenchuk and P.M.Santini, 
%{\it Partially integrable systems in multidimensions by a variant of
%the dressing method: part 1},
J. Phys. A: Math. Gen. {\bf 39} (2006)
5825
%-5845\\



\bibitem{BLR}
M. Bruschi, D. Levi  and O. Ragnisco,   Nuovo Cimento B {\bf 74} (1983) 33


\bibitem{Zen}
 A.I. Zenchuk, Phys.Lett.A {\bf 375} (2011) 2704-2713
%{\it A modification of the method of characteristics: A new class of multidimensional
%partially integrable nonlinear systems}




\end{thebibliography}
\end{document}